\documentclass{article}

\usepackage{arxiv}

\usepackage[utf8]{inputenc} 
\usepackage[T1]{fontenc}    
\usepackage{hyperref}       
\usepackage{url}            
\usepackage{booktabs}       
\usepackage{amsfonts}       
\usepackage{nicefrac}       
\usepackage{microtype}      
\usepackage{lipsum}		
\usepackage{graphicx}
\usepackage{natbib}
\usepackage{doi}
\usepackage{siunitx}
\DeclareSIUnit\bar{bar}
\usepackage{subfig}
\usepackage{xspace}
\usepackage{soul}
\usepackage{amsmath}

\newcommand{\valint}{Val\(_\mathrm{int}\)\xspace}
\newcommand{\valedge}{Val\(_\mathrm{edge}\)\xspace}
\newcommand{\valext}{Val\(_\mathrm{ext}\)\xspace}
\newcommand{\sptbro}{STP-NRTL\(_\mathrm{Bro}\)\xspace }
\newcommand{\sptddb}{STP-NRTL\(_\mathrm{DDB}\)\xspace }
\newcommand{\sptfull}{STP-NRTL\(_\mathrm{Full}\)\xspace }

\title{SPT-NRTL: A physics-guided machine learning model to predict thermodynamically consistent activity coefficients }

\author{ \href{https://orcid.org/0000-0001-5764-9757}{\includegraphics[scale=0.06]{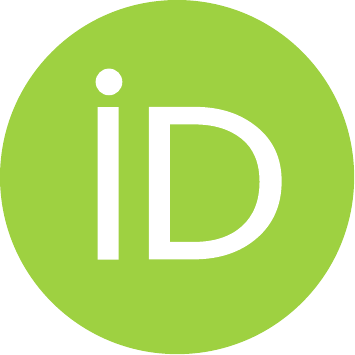}\hspace{1mm}Benedikt A.~Winter}\\
	Energy and Process System Engineering\\
	ETH Zürich\\
	Tannenstrasse 3, 8092 \\
	Zürich, Switzerland \\
	\texttt{bewinter@ethz.ch} \\
	\And
	{\hspace{1mm}Clemens S.~Winter} \\
	OpenAI\\
	3180 18TH St, CA 94110\\
	San Francisco, USA\\
	\texttt{clemenswinter1@gmail.com} \\
	\And 
	\href{https://orcid.org/0000-0002-2552-4391}{\includegraphics[scale=0.06]{orcid.pdf}\hspace{1mm}Timm~Esper}\\
	ITT\\
	University of Stuttgart\\
	Pfaffenwaldring 9, 70569 \\
	Stuttgart, Germany \\
	\texttt{esper@itt.uni-stuttgart.de} \\
	\And 
	\href{https://orcid.org/0000-0001-8013-5439}{\includegraphics[scale=0.06]{orcid.pdf}\hspace{1mm}Johannes ~Schilling}\\
	Energy and Process System Engineering\\
	ETH Zürich\\
	Tannenstrasse 3, 8092 \\
	Zürich, Switzerland \\
	\texttt{jschilling@ethz.ch} \\
	\And 
	\href{https://orcid.org/ 0000-0002-3831-0691}{\includegraphics[scale=0.06]{orcid.pdf}\hspace{1mm}Andr\'e ~Bardow}\thanks{Corresponding author} \\
	Energy and Process System Engineering\\
	ETH Zürich\\
	Tannenstrasse 3, 8092 \\
	Zürich, Switzerland \\
	\texttt{abardow@ethz.ch} \\
}

\renewcommand{\shorttitle}{\textit{arXiv} Template}

\hypersetup{
pdftitle={A template for the arxiv style},
pdfsubject={q-bio.NC, q-bio.QM},
pdfauthor={David S.~Hippocampus, Elias D.~Striatum},
pdfkeywords={First keyword, Second keyword, More},
}

\begin{document}
\maketitle

\begin{abstract}
    The availability of property data is one of the major bottlenecks in the development of chemical processes, often requiring time-consuming and expensive experiments or limiting the design space to a small number of known molecules. This bottleneck has been the motivation behind the continuing development of predictive property models. For the property prediction of novel molecules, group contribution methods have been groundbreaking. In recent times, machine learning  has joined the more established property prediction models. However, even with recent successes, the integration of physical constraints into machine learning models remains challenging. Physical constraints are vital to many thermodynamic properties, such as the Gibbs-Duhem relation, introducing an additional layer of complexity into the prediction. Here, we introduce SPT-NRTL, a machine learning model to predict thermodynamically consistent activity coefficients and provide NRTL parameters for easy use in process simulations. The results show that SPT-NRTL achieves higher accuracy than UNIFAC in the prediction of activity coefficients across all functional groups and is able to predict many vapor-liquid-equilibria with near experimental accuracy, as illustrated for the exemplary mixtures water/ethanol and chloroform/n-hexane. To ease the application of SPT-NRTL, NRTL-parameters of \num{100000000} mixtures are calculated with SPT-NRTL and provided online.     
\end{abstract}


\section{Introduction}

The sheer limitless number of molecules offers a nearly borderless molecular space for chemical engineering in which to find new processing materials like solvents and products. However, limited time and experimental resources make the experimental exploration of this molecular space unfeasible. Thus, the goal to explore the molecular space \textit{in silico} has been driving the development of molecular property prediction. Over the years, the research on predicting molecular properties has led to many approaches based on, e.g., QSPRs \citep{Katritzky.1995,Hughes.2008}, quantum mechanics \citep{Klamt.1995,Lin.2002,Schleder.2019} and group contribution (GC) models \citep{Fredenslund.1975,Marrero.2001,Hukkerikar.2012, Sauer.2014}. More recently, machine learning approaches have been proposed to predict molecular properties \citep{Liu.2019,Ding.2021,Venkatasubramanian.2019,Alshehri.2021}, which, depending on their architecture, can be interpreted as advanced group contribution methods.

Among the many molecular properties of interest to chemical engineering, activity coefficients stand out. Activity coefficients  govern the phase equilibria in essential chemical processes, including distillation and extraction. Since activity coefficients are mixture properties, the combinatorial complexity of mixtures makes the experimental exploration of all activity coefficients of interest challenging.

Historically, group contribution models like UNIFAC \citep{Fredenslund.1975} or quantum chemical models like COSMO-RS or COSMO-SAC \citep{Klamt.1995,Lin.2002} have demonstrated promising results for predicting activity coefficients. These models allow for the exploration of large molecular spaces for which no experimental data is available \citep{Scheffczyk.2016}. However, due to the nature of group contribution methods, UNIFAC cannot predict properties of molecules consisting of non-parameterized groups. In contrast, quantum chemical models like COSMO enable property predication for arbitrary molecules but require computationally demanding calculations for each new molecule.

To overcome these limitations of GC and quantum chemical models and improve overall accuracy, several machine learning approaches have been proposed in recent years for predicting limiting activity coefficients. These machine learning approaches are based on matrix completion \citep{Jirasek.2020,Damay.2021}, collaborative filtering \citep{Tan.2022}, and graph neural networks \citep{SanchezMedina.2022,Rittig.6232022}, and generally surpass UNIFAC and COSMO-RS in accuracy, thereby highlighting the potential of machine learning models for property prediction. 

Recently, we introduced the so-called SMILEStoPropertiesTransformer (SPT), a machine learning model for the prediction of limiting activity coefficients from arbitrary SMILES based on natural language processing \citep{Winter.15.06.2022}. SPT almost halves the mean absolute error compared to UNIFAC and COSMO-RS in the prediction of limiting activity coefficients. Moreover, compared to other machine learning approaches, higher accuracies are obtained in predicting activity coefficients of completely unknown mixtures. Due to the overall promising performance and adaptability of recent machine learning approaches, machine learning is likely to become an integral part of the chemical engineering toolbox for property prediction. \citep{Dobbelaere.2021}

However, while the recent advance in predicting limiting activity coefficients using machine learning models is promising, the use of such models is limited as most applications in chemical engineering require concentration-dependent activity coefficients to calculate phase equilibria. Thus, for a broad application in chemical engineering, machine learning models for predicting activity coefficients have to satisfy three requirements: First, the model must predict concentration-dependent activity coefficients at higher accuracy than state-of-the-art models such as UNIFAC and COSMO-RS/SAC. Second, the model has to be thermodynamically consistent. Third, the model has to be easily applicable in standard process simulation software that engineers use on a daily basis to achieve a high level of adoption.

The concentration-dependency and high accuracy addressed by the first requirement can likely be satisfied by training the model with sufficiently large databases of concentration-dependent activity coefficients such as the Dortmund Datenbank \citep{DortmundDatenbank.2022}. In contrast, thermodynamic consistency and broad application introduce more fundamental challenges to machine learning models. Generally, a basic machine learning model does not guarantee thermodynamic consistency or fulfilling physical constraints because no physical knowledge is incorporated. For concentration-dependent activity coefficients, the Gibbs-Duhem equation relates the derivatives of the activity coefficients to each other. To ensure thermodynamic consistency, the Gibbs-Duhem equation has to be implied by the activity coefficient model, e.g., by expressing the activity coefficients as derivatives of the excess Gibbs energy. Integrating physical knowledge into machine learning models is an active research area with multiple approaches developed  \citep{Rajulapati.2022}, such as predicting parameters of known physical equations \citep{Swischuk.2019} or introducing constraints into the loss function or the neural network itself \citep{beucler.2021}.   

The third requirement concerns the generic applicability of the model in standard engineering tools such as process simulation software. While individual, in-house process models can likely adapt their property methods, the straightforward integration of machine learning models into more commercialized process simulation software by the user is usually only possible to a limited extent. Thus, machine learning models that predict not only thermodynamic properties but also parameters of established models are desirable since they can be directly used in standard process simulation software.

Concerning the prediction of concentration-dependent activity coefficients, \citet{Felton.2022} recently proposed a graph neural network to predict parameters of a fourth-order polynomial fitted to activity coefficients calculated with COSMO-RS. The machine learning model of \citet{Felton.2022} allows the prediction of concentration-dependent activity coefficients with little deviation from COSMO-RS but at much higher speeds. However, the fourth-order polynomial does not ensure thermodynamic consistency, nor does it enable straightforward and easy integration into standard process simulation software. Moreover, as the model is solely trained to COSMO-RS data, the accuracy of the model is expected to be at most that of COSMO-RS. To our knowledge, no machine learning model exists so far for concentration-dependent activity coefficients in the literature that satisfies all of the requirements established above.

In this work, we present a machine learning model to predict thermodynamically consistent, concentration-dependent activity coefficients of binary mixtures from the SMILES representation of molecules. For this purpose, we extend the SPT model recently proposed by the authors for the prediction of limiting activity coefficients \citep{Winter.15.06.2022} by embedding the NRTL equation in the head of the model. Due to the embedded NRTL equation, the resulting so-called SPT-NRTL model provides inherently Gibbs-Duhem consistent activity coefficients. Moreover, SPT-NRTL predicts the NRTL-Parameters internally that can be readily used in many standard process simulation software, including Aspen Plus \citep{AspenTech.2022}, gPROMS \citep{PSE.2022}, or AVEVA \citep{AVEVA.2022}, allowing for an easy application of the model results in existing chemical engineering software.

\section{The SPT-NRTL model architecture}

\begin{figure}[htbp]
    \centering
    \includegraphics[width=0.8\textwidth]{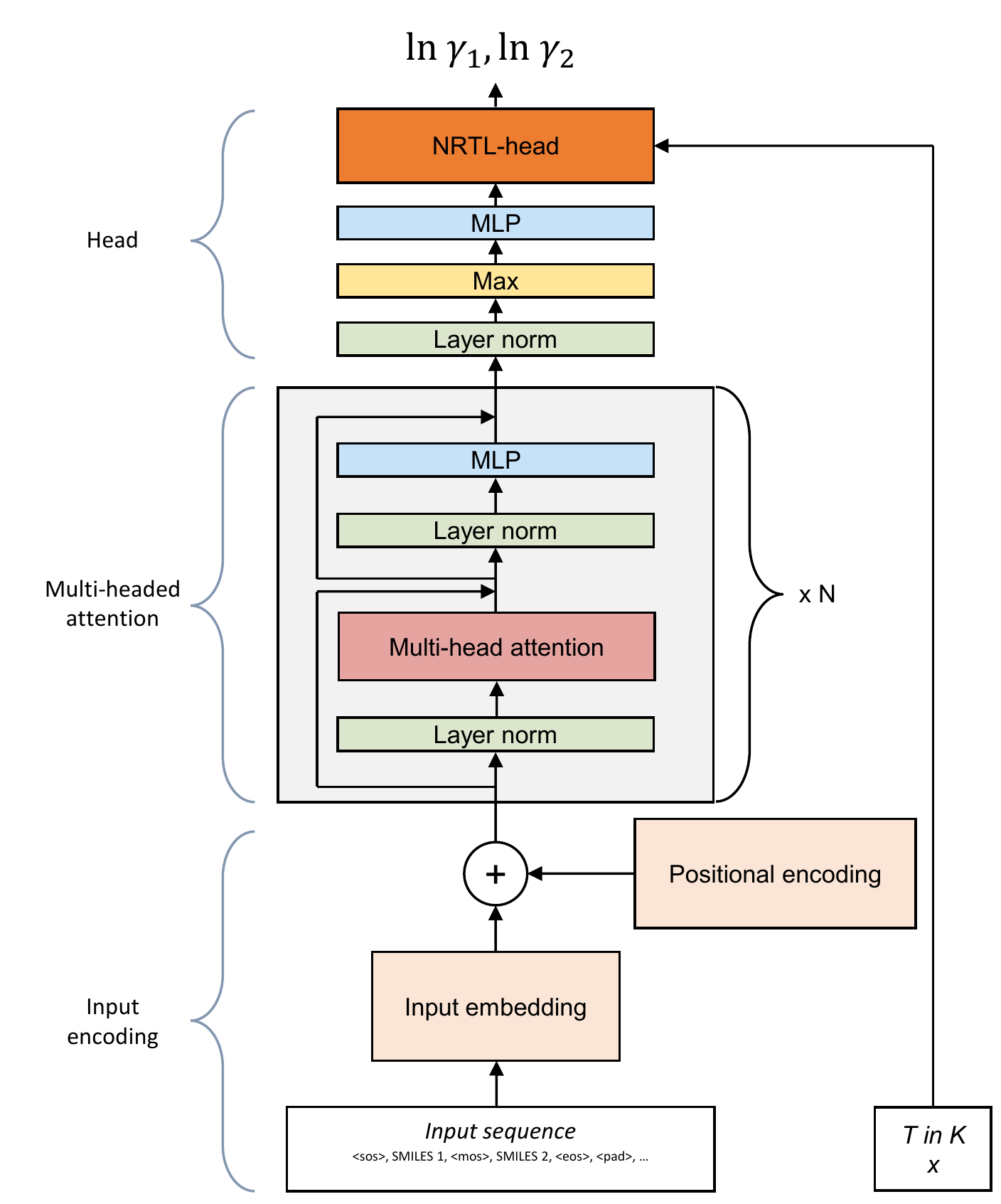}
    \caption{Architecture of SPT-NRTL to predict binary activity coefficients from SMILES codes. The model takes the input sequence consisting of the SMILES of the solvent and solute as input. In the input encoding section of the model, the information about the individual tokens making up the SMILES and the position of tokens are compiled into a single matrix. The multi-headed attention section transmits information between the parts of the molecules. The head section of SPT-NRTL first reduces the high dimensional output of the transformer. Afterward, the output is passed to the NRTL-Head, which contains the NRTL equation. The NRTL-Head takes the mole fraction x and temperature T as additional inputs. 
}
    \label{fig:network}
\end{figure}

The SPT-NRTL model should enable the calculation of thermodynamically consistent, concentration-dependent activity coefficients of binary mixtures using natural language processing. In this section, the model architecture of SPT-NRTL is introduced (Figure~\ref{fig:network}). The architecture of SPT-NRTL is based on the architecture of the natural language model GPT-3 \citep{Brown.28.05.2020} using a decoder-only transformer architecture developed by \citet{Vaswani.12.06.2017}. The transformer architecture has proven suitable for understanding not only the grammar of natural language but also the molecular grammar embedded within SMILES codes, a linear text-based molecular representation introduced by \citet{Weininger.1988}, leading to many successful applications in the field of chemistry \citep{Schwaller.2019,Honda.12.11.2019,Lim.2021,Kim.2021}. 

In the following, we present the SPT-NRTL architecture broken up into three sections: input embedding (Section~\ref{sec:input_embedding}), multi-headed attention (Section~\ref{sec:MHA}), and head (Section~\ref{sec:head}).

\subsection{Input embedding}\label{sec:input_embedding}

SPT-NRTL predicts activity coefficients based on the SMILES codes of the input components. SMILES codes have emerged as one of the standard molecular representations for machine learning in chemical engineering with many recent applications \citep{Honda.12.11.2019,Wang.2019,Schwaller.2019,Lim.2021}. The SMILES code developed by \citet{Weininger.1988} allows for a linear string representation of complex branched and cyclic molecules. In SMILES codes, atoms are encoded as periodic table symbols, e.g., the character "N" for nitrogen, except for hydrogen atoms that are implicitly assumed. Single bonds are implicitly assumed, while double or triple bonds are represented by the characters "=" and "\#", respectively. Branches are contained within brackets, and joints of ring structures are represented as numbers. Thus, for example, the molecule 2-ethyl phenol can be represented by the following SMILES: Oc1c(CC)cccc1.

The input of SPT-NRTL consists of the SMILES codes representing the molecules in the mixture concatenated with special characters denoting the start of the sequence $<\mathrm{SOS}>$, the separation between the molecules $<\mathrm{MOS}>$, and the end of the last molecule $<\mathrm{EOS}>$. The remainder of the input sequence is filled up to a sequence length $n_\mathrm{seq}$ of \num{128} with padding $<\mathrm{PAD}>$: 
 \begin{equation}
\centering
    <\mathrm{SOS}>,\mathrm{SMILES_{0}},<\mathrm{MOS}>,\mathrm{SMILES_{1}},<\mathrm{EOS}>,<\mathrm{PAD}>,... \nonumber
    \label{eq:input}
\end{equation}

To make the input string readable for the machine learning model, the input string is tokenized, meaning that the sequence is broken up into tokens that can each be represented by a single number. In general, tokens can consist of multiple characters. However, in this work, every token consists of one character and is assigned a unique number. Water, represented by the SMILES 'O', is assigned its own token. The tokenization process of the SMILES can be seen as analogous to assigning first-order groups in group contribution methods. The full vocab containing all tokens is available in the Supporting Information Section 1.

The input sequence is then encoded using one-hot encoding, where each token is represented by a learned vector of size $n_\mathrm{emb} = 512$. An input matrix of size $n_\mathrm{emb} \times n_\mathrm{seq}$ is constructed by concatenating the vectors representing the tokens in the input sequence. After the input sequence is encoded, an additional vector is concatenated to the right of the input matrix, which can contain a linear projection of continuous variables into embedding space. In the case of the original SPT model~\citep{Winter.15.06.2022}, temperature information is encoded in this vector. However, other continuous variables, such as pressure, mole fraction, etc., could also be included. In SPT-NRTL, no continuous variables are supplied here since temperature and mole fraction information enters the model only in the last stage (cf., Figure~\ref{fig:network}).  After adding the continuous variables, the resulting input matrix has a size of $n_\mathrm{emb} \times n_\mathrm{seq+1}$. Next, a learned positional encoding of size $n_\mathrm{emb} \times n_\mathrm{seq+1}$ is added to the input matrix. At this point, the input matrix contains information on all atoms and bonds in the molecular structure of the mixture components as well as their position. However, each token has no information about its surrounding as no information has been shared between tokens yet. This information sharing between tokens will take place in the next section, the multi-headed attention.

\subsection{Multi-headed attention}\label{sec:MHA}

In the multi-headed attention section, multiple multi-headed attention blocks are stacked on top of each other.  Within each block, the input is first normalized by a layer norm and then passed to the multi-headed attention mechanism. Within the multi-headed attention mechanism, information is transferred between tokens. While individual tokens only contain information about themselves after the input encoding, the multi-headed attention mechanism allows tokens to obtain information about their neighbors or other atoms of interest within their own molecule or even other molecules in the mixture. Thus, a transformer block could be interpreted as a self-learning n\textsuperscript{th}-order group contribution method, where each token, or smallest possible group, learns the importance of other tokens and then self-assembles higher-order groups based on the structure of the molecule.

On a mathematical level, the output $Z_i$ of a single attention head $i$ is defined as:

\begin{equation}
\label{equ:att}
Z_i = \mathrm{softmax} \left( \frac{Q_iK_i^\mathrm{T}}{\sqrt{d_k}} \right) V_i
\end{equation}
with the query matrix $Q_{i}$, the key matrix $K_{i}$, the value matrix $V_{i}$, and $d_\mathrm{k} = d_\mathrm{emb} / n_\mathrm{head}$, where $n_\mathrm{head}$ is the number of attention heads. The output $Z_i$ of each head $i$ is concatenated and projected to the size $n_\mathrm{emb} \times n_\mathrm{seq+1}$, which is passed to a multilayer perceptron (MLP) that ends the transformer block. For a more in-depth and visual explanation, the reader is referred to the blog of \cite{Alammar.2018} or our previous work \citep{Winter.15.06.2022}.

\subsection{Head}\label{sec:head}

\begin{figure}[htbp]
    \centering
    \includegraphics[width=0.5\textwidth]{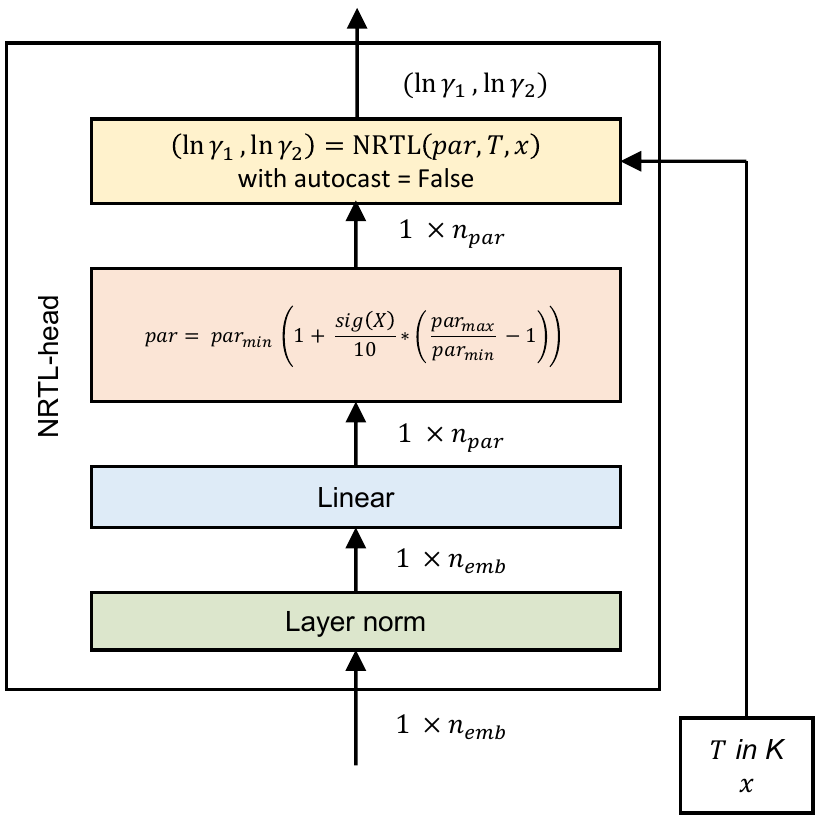}    
    \caption{Architecture of the NRTL-head. The input vector to the NRTL-head of size $1\times \mathrm{emb}$ is passed through a layer norm and then projected to the number of parameters $n_\mathrm{par}$ of the NRTL equation. The NRTL parameters are calculated from lower and upper bounds with Eq. \ref{eq:param}. Finally, $\ln\gamma_1$ and $\ln\gamma_2$ are calculated using the NRTL equation (Eq. \ref{eq:NRTL})}
    \label{fig:head}
\end{figure}

In the head section, the output dimension of the multi-headed attention block is reduced, and physical constraints are applied. First, a layer norm followed by a max function is applied along the sequence dimension, which reduces the output to size $n_\mathrm{emb} \times 1$. The result  passes to a multilayer perceptron (MLP) and afterward into the NRTL-head. The architecture of the NRTL-head is shown in Figure~\ref{fig:head}. Introducing the exponential NRTL function in the head of the machine learning model introduces instability to the training of the model, often leading to divergence of the model during training. Thus, multiple measures are included in the NRTL-head to improve stability.

Within the NRTL-head, the input of the NRTL-head is first normalized in a layer norm and then projected linearly from $n_\mathrm{emb}$ to the number of NRTL parameters $n_\mathrm{par} = 10$ ($a_{1\text{-}2}$, $t_{12,1\text{-}4}$,$t_{21,1\text{-}4}$). Afterward, the NRTL parameters are calculated from the output $X$ of the linear layer as follows, using a sigmoid function and upper and lower bounds for the NRTL parameters:
\begin{equation}
    par = par_\mathrm{min} \left(1 + \frac{\mathrm{sig(X)}}{10} * \left( \frac{par_\mathrm{max}}{par_\mathrm{min}}\right) \right)
    \label{eq:param}
\end{equation}
where, $par_\mathrm{min}$ and $par_\mathrm{max}$ are lower and upper bounds for NRTL parameters, determined from a database of NRTL parameters fitted to a large range of mixtures in a previous study \citep{Scheffczyk.2016}. Constraining the NRTL parameters improves the convergence of the training of the model. For the NRTL parameters $a_1$, $t_{12,1}$ and $t_{21,1}$, $par_\mathrm{min}$ and $par_\mathrm{max}$ are not symmetric. Asymmetric bounds ensure better starting values at the beginning of the training, where all parameters would likely be the mean of the upper and lower bound, and asymmetric bounds thus result in non-zero starting values. Afterward, the obtained NRTL parameters are passed into the NRTL model given in Eq.~\eqref{eq:NRTL} to Eq.~\eqref{eq:NRTL_tau}. We assume a symmetric $\alpha$ in the NRTL equation to be compatible with the implementation in the flowsheeting software Aspen.
\begin{align}\label{eq:NRTL}
     \ln\gamma_1 &= x_2^2 \left[ \tau_{21}\left( \frac{G_{21}}{x_1+x_2 G_{21}}\right)^2 + \frac{\tau_{12}G_{12}}{\left(x_2+x_1 G_{12}\right)^2}  \right] \\
     \ln\gamma_2 &= x_1^2 \left[ \tau_{12}\left( \frac{G_{12}}{x_2+x_1 G_{12}}\right)^2 + \frac{\tau_{21}G_{21}}{\left(x_1+x_2 G_{21}\right)^2}  \right]
\end{align}
with
\begin{align}
    &\ln G_{12} = -\alpha \tau_{12}\\
    &\ln G_{21} = -\alpha \tau_{21}
\end{align}    
and
\begin{align}
     &\alpha = a_0 + a_1 T \\
     &\tau_{i,j} = t_{i,j,0} + \frac{t_{i,j,1}}{T} + t_{i,j,2} \ln T + t_{i,j,3} T \label{eq:NRTL_tau}
\end{align}

When calculating the NRTL equation, autocasting of floating point numbers to FP16 is disabled, and FP32 is used instead to improve the stability of the model.

The NRTL-head returns $\ln\gamma_1$ and $\ln\gamma_2$ as the final output of the model. During evaluation, the NRTL parameters are internally calculated and returned as additional output of the model. While the predicted activity coefficients dependent on the input composition and temperature, the predicted NRTL parameters are independent of composition and temperature and are thus suitable for a wider application range. 

\section{Property data for training and validation}

Natural language processing models typically require large amounts of training data to reach high generalization capabilities. For the prediction of physical properties, these large amounts of experimental training data are often not available. Thus, natural language processing models predicting physical property need pretraining. Two options for pretraining have emerged: one based on auto-translation tasks that first teach the models the grammar of SMILES before introducing property data in a subsequent step \citep{Honda.12.11.2019} and another where synthetic data is used \citep{Winter.15.06.2022,Vermeire.2021}. To our knowledge, no study comparing the performance of both approaches exists to date. For SPT-NRTL, we follow the approach of SPT and introduce an initial pretraining step on synthetic property data sampled from COSMO-RS. In the second step, the model is fine-tuned on experimental data. In the following, the synthetic datasets (Section \ref{sec:S_data}) and experimental datasets (Section \ref{sec:E_data}) are described.

\subsection{Synthetic data for pretraining}\label{sec:S_data}

For the pretraining of our SPT-NRTL model, we extend the synthetic dataset on limiting activity coefficients already used for the pretraining of our SPT model \citep{Winter.15.06.2022}. The synthetic dataset is sampled from the COSMObase 2020 database containing around \num{10000} non-ionic components. For SPT, the pretraining dataset contained 5 million data points sampled at infinite dilution and \SI{298.15}{\kelvin}, and 5 million data points sampled at infinite dilution and a random temperature between \SI{278.15}{\kelvin} and \SI{598.15}{\kelvin} (cf., \citet{Winter.15.06.2022}). For SPT-NRTL, we extend this dataset by 3 million data points sampled at a random concentration at \SI{298.15}{\kelvin}. This approach yields a total amount of around 13 million data points. Since water has proven difficult to predict, we additionally sampled mixtures of water and all other components five times at a random concentration and temperature between \SI{278.15}{\kelvin} and \SI{598.15}{\kelvin}. To add additional variation to the data and increase the robustness of the model towards the order of input components, both possible permutations of mixture order are included in the data, yielding 26 million data points in total.

\subsection{Experimental data for fine-tuning}\label{sec:E_data}

Two experimental datasets are available for fine-tuning the model: First, a dataset of experimental limiting activity coefficients based on \citet{Brouwer.2021}, which was previously used for the fine-tuning of the SPT model. The dataset has been cleaned regarding data entry errors and undefined components (for more details, see \citet{Winter.15.06.2022}). The resulting Brouwer (BRO) dataset contains \num{20870} data points with \num{349} solvents and \num{373} solutes in \num{6416} unique combinations at temperatures ranging from \SI{250}{\kelvin} to \SI{555.6}{\kelvin}.

The second available dataset is calculated from the vapor-liquid equilibrium (VLE) and vapor pressure (VAP) datasets of the Dortmund Database (DDB) \citep{DortmundDatenbank.2022}. We consider only mixtures with a pressure below \SI{5}{\bar} and exclude mixtures of molten metals. Overall, the resulting DDB dataset of concentration-dependent activity coefficients contains \num{77053} data points with \num{506} components in \num{2302} unique combinations at temperatures between \SI{183.0}{\kelvin} and \SI{973.15}{\kelvin}. For the considered pressure range, we assume an ideal gas phase and neglect the Poynting factor. Thus, concentration-dependent activity coefficients are calculated for the DDB dataset as: 
\begin{equation}
    \gamma_i = \frac{py_i}{p_\mathrm{sat,i} x_i}
    \label{eq:ddb}
\end{equation}
where $p$ is the pressure, $y_i$ is the mole fraction of component $i$ in the gas phase, $p_\mathrm{sat,i}$ is the vapour pressure of component $i$ and $x_i$ is the mole fraction of the component $i$ in the liquid phase. 

The distribution of the training data is shown in Supporting Information Section 2.

To evaluate the performance of our SPT-NRTL model, three training data\-sets are created from the Brouwer and DDB datasets:

\begin{enumerate}
    \item	Brouwer: Containing only the Brouwer dataset and thus only limiting activity coefficients
    \item	DDB: Containing only the DDB dataset and thus only concentration-dependent activity coefficients without limiting activity coefficients
    \item	Full: Containing both the Brouwer and DDB dataset and thus limiting and concentration-dependent activity coefficients
\end{enumerate}

These three datasets allow us to assess whether large commercial datasets are required to train accurate models or if smaller open-source datasets of solely limiting activity coefficients are sufficient.

For the validation of the model, we follow the strategy of the previous SPT model by defining three validation sets \citep{Winter.15.06.2022}:
\begin{enumerate}
    \item Val$_\mathrm{int}$: Contains binary mixtures where both components are contained in the training set but not in this exact combination. This validation set allows us to validate the interpolation capabilities of the model for well-known regions where experimental data is available for both components, however, not for this exact combination of components. 
    \item Val$_\mathrm{edge}$: Contains binary mixtures where either the solvent or the solute are contained in the training set but not both. This validation set allows us to validate extrapolation capabilities for regions where only one component is known, for example, when a new solvent has to be identified for a known solute.
    \item Val$_\mathrm{ext}$: Contains binary mixtures where neither the solvent nor the solute are contained in the training set. This validation set allows us to validate the extrapolation capabilities for regions where new components are investigated.
\end{enumerate}
Overall, these three validation sets thus allow to test both the interpolation and extrapolation capabilities  of the SPT-NRTL model.

{To create the three validation sets, we use n-fold cross-validation: First, the unique mixtures of all combined datasets are split into $n$ Val$_\mathrm{ext}$ datasets. Then, for each dataset $i$ of the $n$ Val$_\mathrm{ext}$ datasets, mixtures not contained in Val$_\mathrm{ext,i}$, but containing one component contained within a mixture in Val$_\mathrm{ext,i}$, are sorted into Val$_\mathrm{edge,i}$.}

Subsequently, mixtures contained in the validation sets Val$_\mathrm{ext,i}$ and Val$_\mathrm{edge,i}$ are removed from the training datasets, i.e., Train$_\mathrm{BRO,i}$, Train$_\mathrm{DDB,i}$, and Train$_\mathrm{full,i}$. Afterward, \SI{5}{\percent} of the mixtures remaining in the training sets are sampled into Val$_\mathrm{int,i}$, and all systems are reassessed whether they have to be moved into another validation set due to the removal of Val$_\mathrm{int,i}$. The resulting datasets vary in size between \num{9}-\num{733} for \valext, \num{9}-\num{733} for \valedge and \num{9}-\num{733} for \valint due to the different frequency of measurements between mixtures. Since all data is used to define the validation sets, the SPT-NRTL model is validated on similar datasets containing data from both the DDB dataset and Brouwer dataset, independent of the dataset used for fine-tuning. This consistent validation set allows for a consistent comparison between the later resulting models.

\section{Training of SPT-NRTL}\label{sec:training}

Starting training of SPT-NRTL from a freshly initialized model showed poor convergence and only converged with low learning rates (\num{1e-6}), which results in a high final loss. To overcome the training instabilities, we, therefore, start the training of the SPT-NRTL model not from a freshly initialized model but from the SPT model pre-trained on limited activity coefficients from COSMO-RS \citep{Winter.15.06.2022} where the head section of the model is exchanged from a regression head to an NRTL-head. Thereby, the model already has the encoding and multi-headed attention section of the model trained when it is introduced to concentration-dependent data. This use of a pretrained model for the pretraining improves stability during training, lowers the amount of required training time, lowers the final loss, and enables higher learning rates.

We first train the SPT-NRTL model, initialized by the SPT model, on the 26 million datapoints of synthetic COSMO dataset for 10 epochs. The resulting model is fine-tuned on experimental data from the Brouwer, DDB, and Full datasets using n-fold cross-validation for additional 50 epochs. The pretraining on the synthetic COSMO dataset requires about \SI{48}{\hour} on a 2080Ti, while the fine-tuning on the experimental data takes between \SIrange{8}{17}{\minute} on a 2080Ti, depending on the dataset.

As SPT-NRTL training is based on an SPT, the hyperparameters of the model architecture remain unchanged from SPT. A detailed breakdown of the hyperparameters is given in the Supporting Information Section 3. During pretraining and fine-tuning, a learning rate of \num{1e-4} and MSE loss is used.

\section{Results: Predicting concentration-depending activity coefficients}

\subsection{Prediction error as function of concentration}

\begin{figure}[htbp]
    \centering
    \captionsetup[subfigure]{oneside,margin={1cm,0cm}}
    \subfloat[Interpolation: \valint]{
    \includegraphics[width=0.3\textwidth]{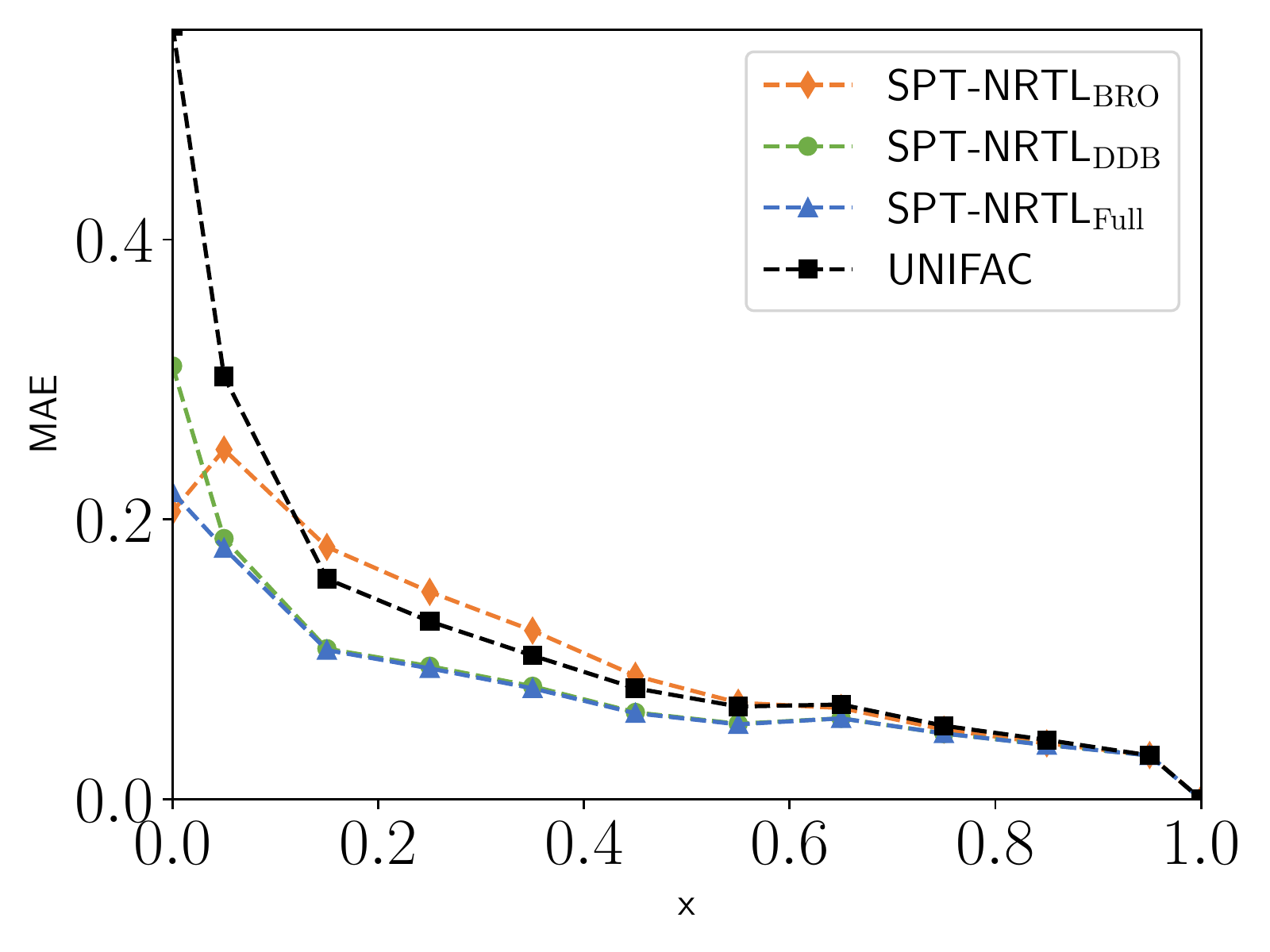}%
    \label{fig:fineint}
    }
    \captionsetup[subfigure]{oneside,margin={1cm,0cm}}
    \subfloat[Edge-Extrapolation: \valedge]{
    \includegraphics[width=0.3\textwidth]{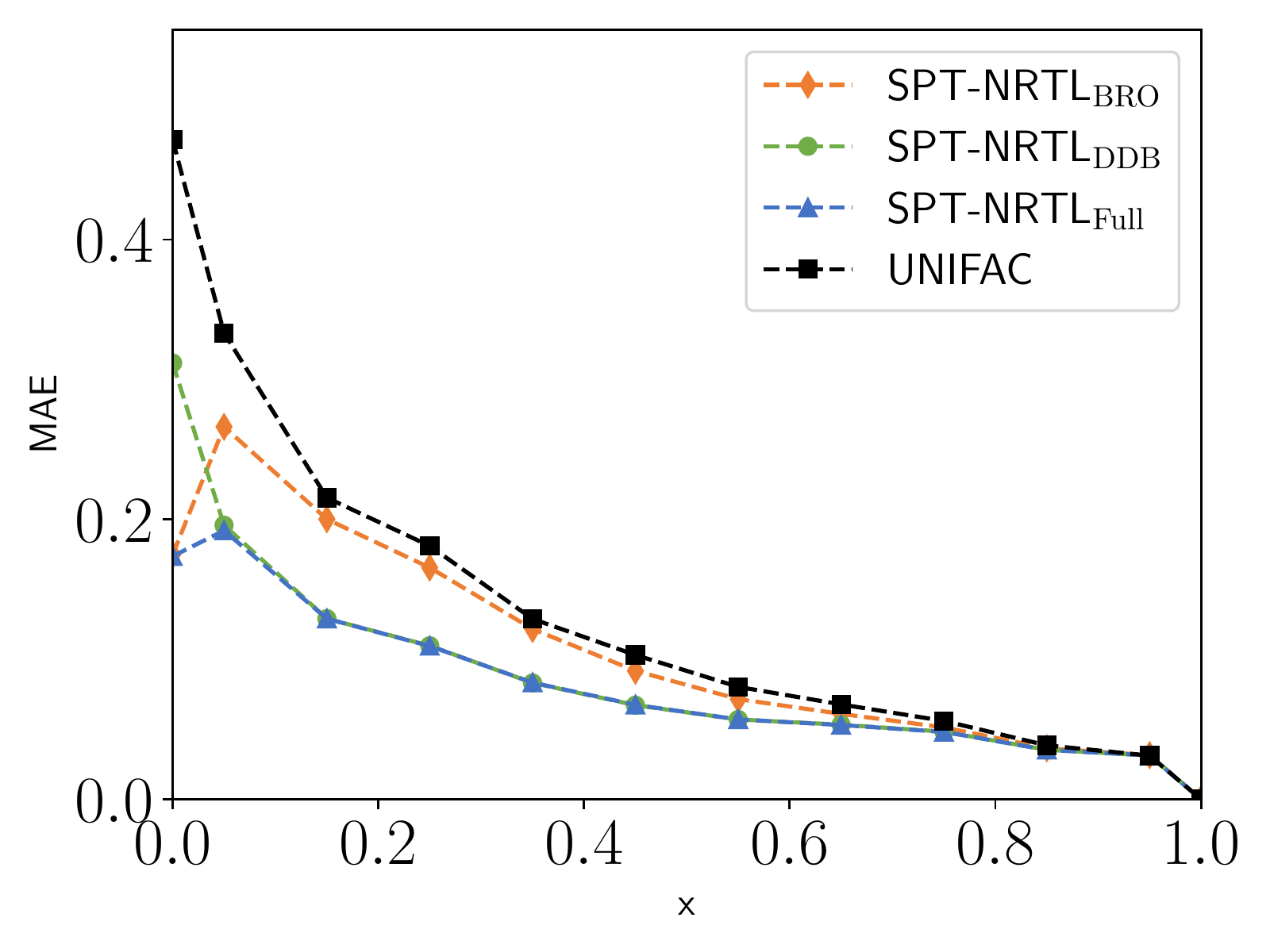}%
    \label{fig:fineedge}
    }
    \captionsetup[subfigure]{oneside,margin={1cm,0cm}}
    \subfloat[Extrapolation: \valext]{
    \includegraphics[width=0.3\textwidth]{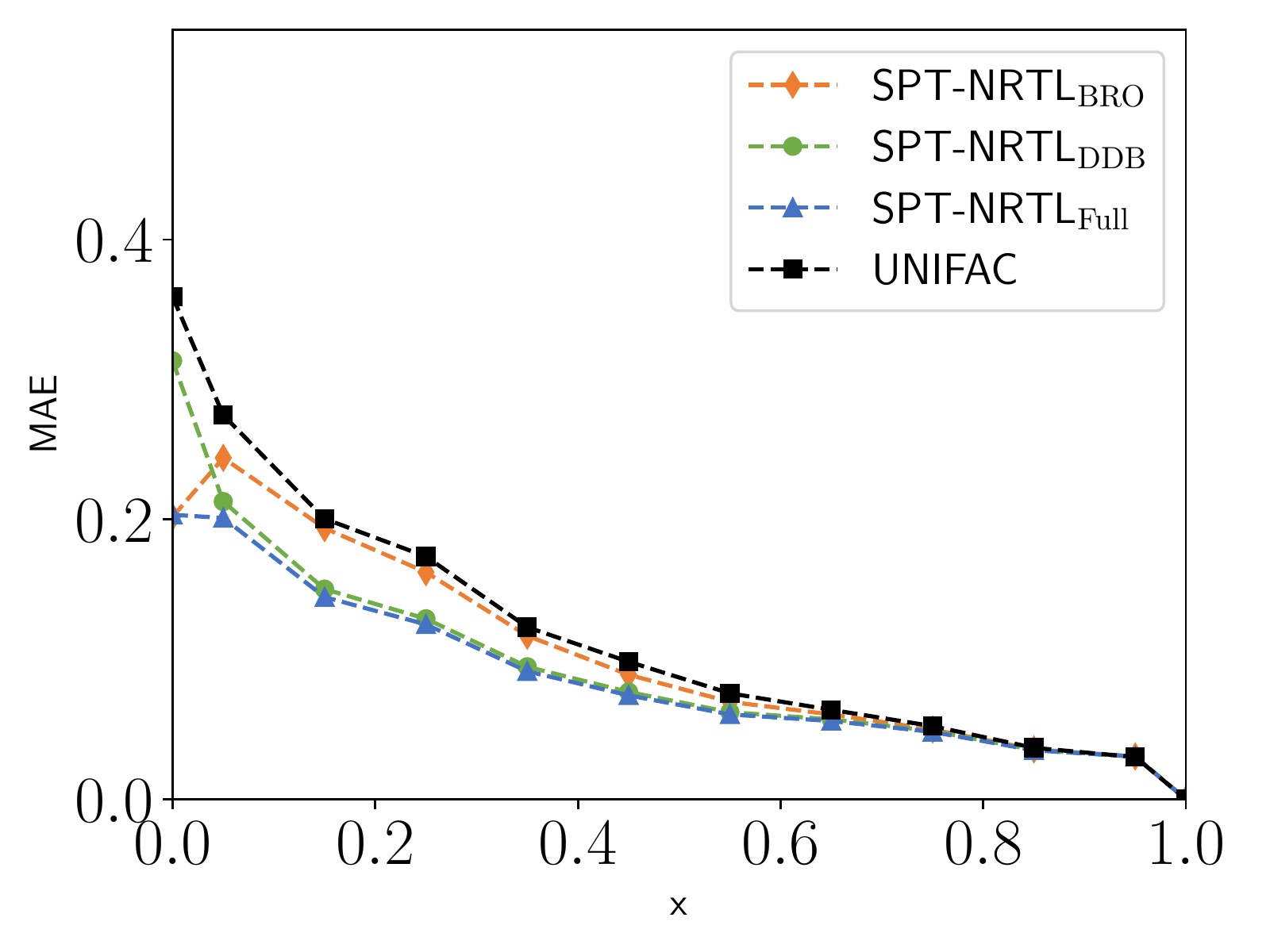}%
    \label{fig:fineext}
    }
    \caption{ Mean absolute error (MAE) of predicted concentration-dependent activity coefficients $\ln \gamma$ for SPT-NRTL fine-tuned on the Brouwer (\sptbro), DDB (\sptddb), and Full (\sptfull) dataset, as well as UNIFAC. The boundary points only include data points with a mole fraction of exactly $x=0$ or $x=1$, while all other points average the error over a concentration interval of $\Delta x =\num{0.1}$}
    \label{fig:fine}
\end{figure}

In the following section, we first present the results of SPT-NRTL fine-tuned on experimental data of the Brouwer, DDB, and Full datasets (cf., Section~\ref{sec:E_data}). These datasets allow us to analyze how the  distribution of the training data concerning concentration affects the performance of the fine-tuned model. The three validation sets are used to assess the interpolation and (edge-) extrapolation capabilities of SPT-NRTL. For this assessment, n-fold cross-validation is used as described in Section~\ref{sec:training}. We compare the performance of predicting con\-cen\-tra\-tion-dependent activity coefficients to the UNIFAC model. For a fair comparison between SPT-NRTL and UNIFAC, we consider only the datapoints that can be calculated by both SPT-NRTL and UNIFAC. Since water is always contained in the training data, \valext does not contain any water mixture resulting in \num{304952} datapoints considered for validation of all SPT-NRTL models and UNIFAC. Generally, water mixtures have a high $\ln \gamma$, and thus, the $\ln \gamma$ is overall lower in \valext because no water mixtures are contained in contrast to \valedge and \valint. This is the reason behind a lower average error of UNIFAC in predicting the mixtures contained in \valext. For \valedge, \num{349796} datapoints are considered for validation of SPT-NRTL and UNIFAC. For \valint, only \num{119116} datapoints are considered for validation because \valint only contains mixtures of components that are part of both the Brouwer and DDB dataset. The mean absolute error (MAE) is used as metric to assess prediction quality. As $\ln \gamma$ increases with decreasing concentration, we analyze the mean absolute error as a function of concentration (see Figure \ref{fig:fine}).

Generally, the mean absolute error of $\ln \gamma$ increases for all models and validation sets with decreasing concentration. This trend results from the general increase of $\ln \gamma$ with decreasing concentration, usually reaching a maximum at infinite dilution ($x=0$).

Overall, \sptfull has the lowest mean absolute error along the overall concentration range for all validation sets. At infinitive dilution, \sptfull has a mean absolute error of \num{0.21} on \valint, \num{0.17} on \valedge and \num{0.2} on \valext compared to a mean absolute error of \num{0.6}, \num{0.47} and \num{0.36} of UNFIAC, respectively. The error of UNIFAC changes with the validation dataset as the predicted mixtures change. The better performance of \sptfull at infinitive dilution compared to the concentration interval of $\Delta x = \left( 0, 0.1 \right]$ in \valedge and \valint results from the unbalanced concentration distribution of the training datasets. The Brouwer dataset contains approximately \SI{50}{\percent} of the data points of the DDB dataset. However, the Brouwer data are all at infinitive dilution, while the data points of the DDB dataset are spread over the concentration range. 

At high concentrations ($x > 0.7$), STP-NRTL performs similarly to UNIFAC on all three validation sets \valint, \valedge and \valext, independent of the dataset used to train SPT-NRTL (Brower, DBB, or Full). For concentrations above $x=0.5$, a mean absolute error below \num{0.1} is achieved for all training and validation datasets. At concentrations below $x=0.5$, SPT-NRTL outperforms UNIFAC in prediction quality, and differences between the three training datasets become more pronounced. 

\sptbro and \sptddb are trained on either data at infinitive dilution or concentration-dependent data without data at infinite dilution, respectively. This imbalance of  training data affects the prediction performance of the SPT-NRTL models: \sptddb performs similarly to \sptfull for most of the concentration range except for the prediction at infinitive dilution ($x=0$), where the mean absolute error significantly increases. This increase at infinitive dilution results from the fact that the DDB dataset used to train \sptddb does not contain data points at infinitive solution. In contrast, \sptbro has a prediction error close to UNIFAC for most of the concentration range except for the prediction at infinitive solution, where the mean absolute error significantly decreases. This decrease results from the fact that the Brouwer dataset used to train \sptbro contains only data at infinitive solution, but no concentration-dependent data. However, although only experimental data at infinitive solution is used for the fine-tuning, the pretraining to synthetic data and the physical basis of the NRTL-head results in a competitive performance of \sptbro to predict concentration-dependent activity coefficients. 

Overall, SPT-NRTL is able to achieve higher accuracy than UNIFAC, particularly at low concentrations ($x < 0.5$). Further improvements in the prediction quality of SPT-NRTL could be achieved by using more curated training data.

\subsection{Performance of molecular families}

In the previous section, the correlation between the mean absolute error and the concentration of a component is investigated for UNIFAC and the SPT-NRTL models \sptbro, \sptddb, and \sptfull. However, prediction errors of molecular properties generally depend on the investigated molecular family (e.g., alcohols or alkanes) and thus on the functional groups present within the molecular structure of the molecule. \citep{Brouwer.2019} In this section, we therefore analyze the prediction error of \sptfull and UNIFAC depending on the functional groups contained in the molecular structure of the solvent and solute.

We focus the analysis on the following molecular families: alcohols, aldehydes, aliphatics, aromatics, carboxylic acids, ethers, halogens, ketones, and nitrates. Moreover, we consider water as additional molecular family. To isolate effects to single functional groups, we consider only mixtures of molecules that contain no more than one analyzed functional group in the molecular structure. All molecules with more than one corresponding functional group or functional groups not captured by the investigated molecular families are discarded for the analysis. 
The remaining molecules are assigned to one of the ten molecular families. A list of all molecules and their assigned groups is available in the Supporting Information Section 5.

\begin{figure}[]
    \centering
    \captionsetup[subfigure]{oneside,margin={1cm,0cm}}
    \subfloat[\sptfull \valext]{
    \includegraphics[width=0.45\textwidth]{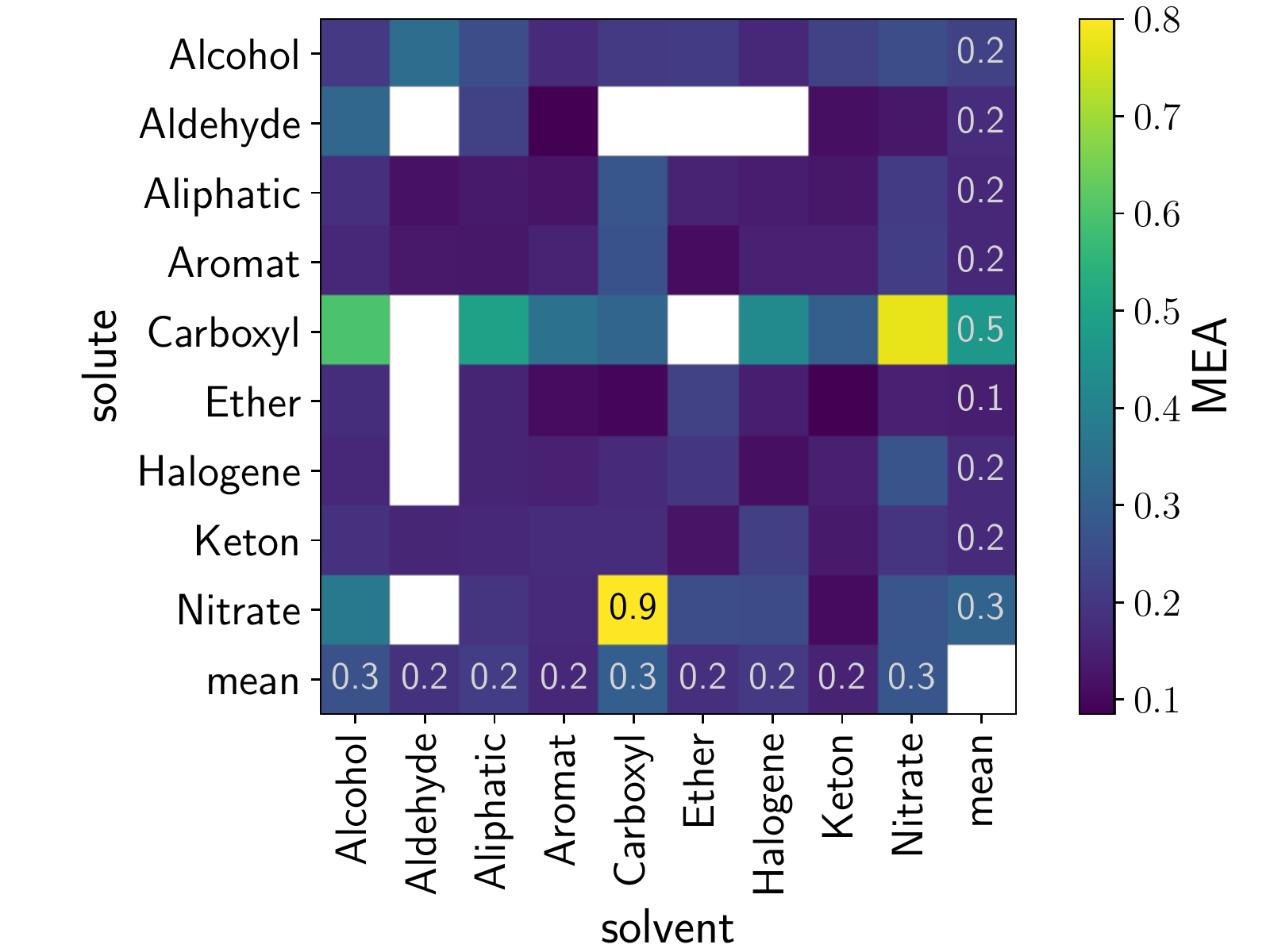}%
    \label{fig:heatext}
    }
    \centering
    \captionsetup[subfigure]{oneside,margin={1cm,0cm}}
    \subfloat[\sptfull \valedge]{
    \includegraphics[width=0.45\textwidth]{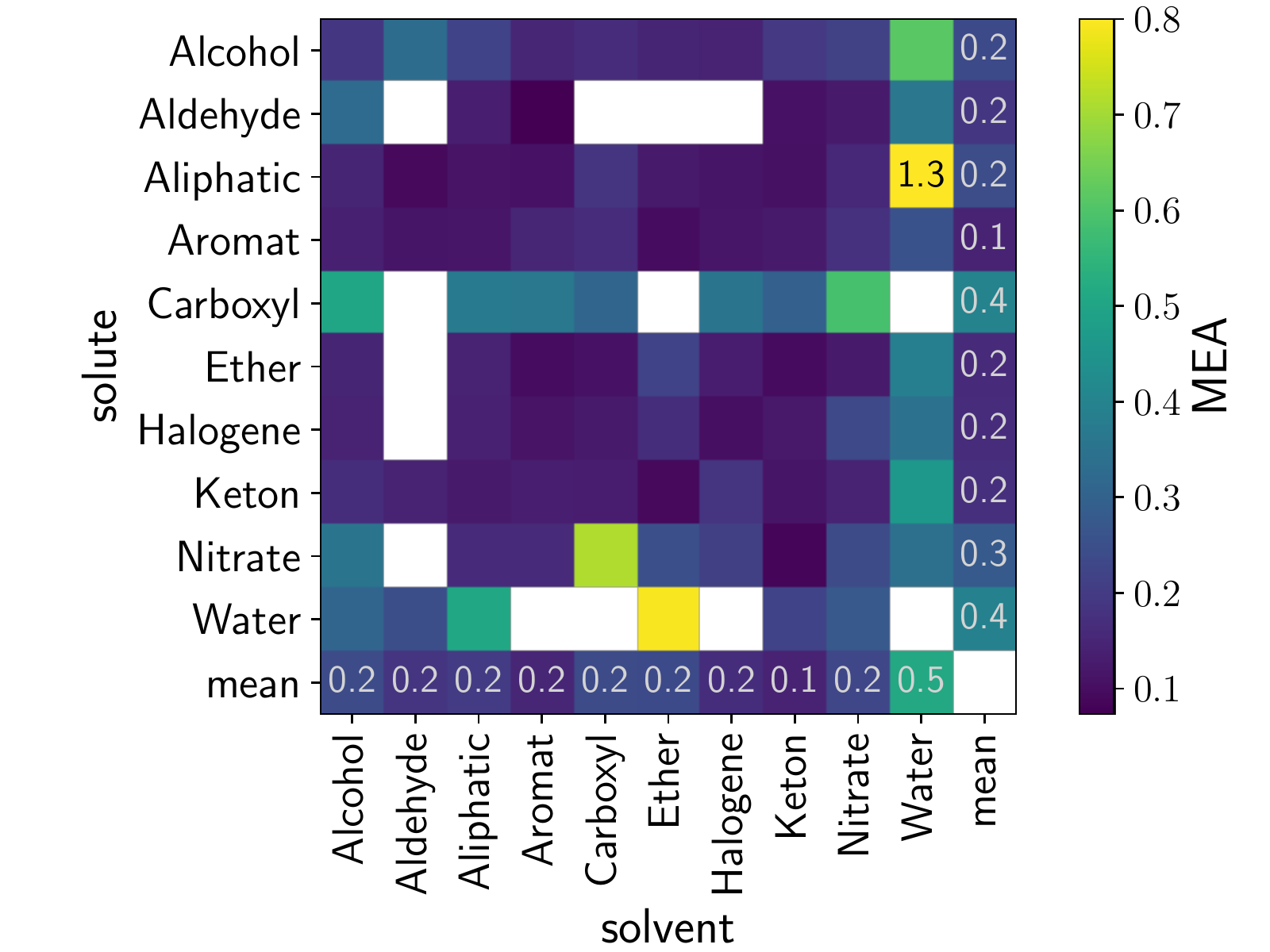}%
    \label{fig:heatedge}

    }
    \centering
    \captionsetup[subfigure]{oneside,margin={1cm,0cm}}
    \subfloat[\sptfull \valint]{
    \includegraphics[width=0.45\textwidth]{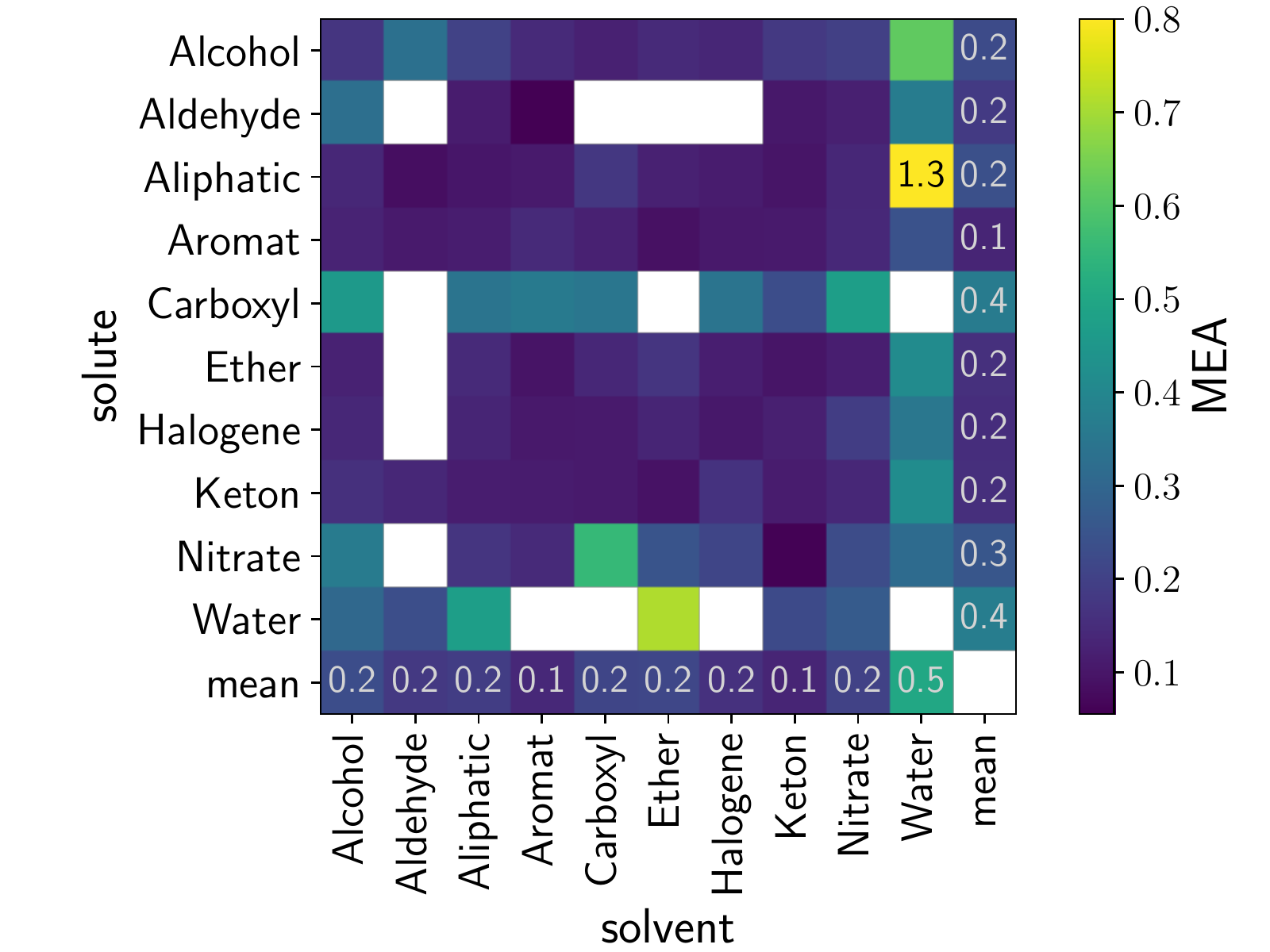}%
    \label{fig:heatint}
    }
    \centering
    \captionsetup[subfigure]{oneside,margin={1cm,0cm}}
    \subfloat[UNIFAC]{
    \includegraphics[width=0.45\textwidth]{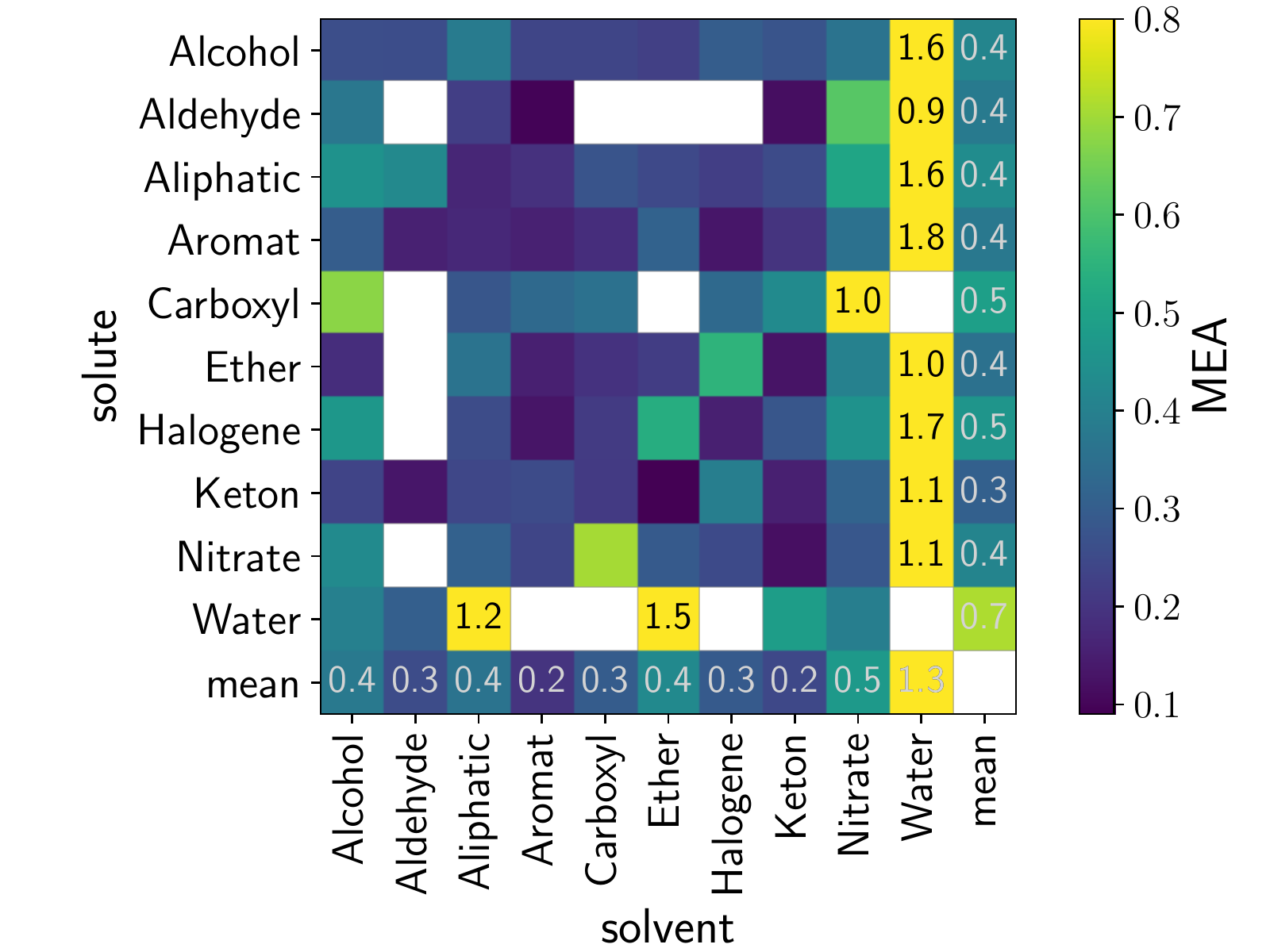}%
    \label{fig:heatuni}
    }
    \caption{Heat map of the mean absolute error  of the activity coefficients for $x_\mathrm{solute} < 0.3$ for \sptfull and UNIFAC depending on the functional group of the solvent and solute. The heat map has a maximum of \num{0.8} for illustration purposes. Mean absolute errors greater than \num{0.8} are explicitly given. The mean column gives the average mean absolute error over all groups.}
    \label{fig:heatval}
\end{figure}

Since the prediction error of $\ln \gamma$ increases with decreasing concentration, only $\ln \gamma$ data points with $x < 0.3$ are considered in this analysis to capture the region of poor performance. The component $i$ with $x_i < 0.3$ is considered the solute, the component $j$ with $x_j = 1-x_i > 0.7$ the solvent. Only the activity coefficient of the solute is considered for the analysis. To calculate the mean absolute error for certain functional group combinations, we consider all mixtures shared between UNFIAC and \sptfull \valint. First, the average of the mean absolute error for each individual mixture in the dataset is calculated to eliminate a bias towards frequently measured mixtures, e.g., water-ethanol. Afterward, the resulting mean absolute errors are averaged within each functional group combination. Combinations with less than five unique mixtures are discarded. (Figure \ref{fig:heatval}).

As a general trend, \sptfull has substantially lower prediction errors than UNIFAC for all validation sets and functional group combinations (Figure \ref{fig:heatval}). For alcohols, aldehydes, aliphatic and aromatic hydrocarbons, ethers, halogens, and ketones, SPT-NRTL achieves a very high prediction quality in the range of \numrange{0.1}{0.2} as long as no water is used as solvent. Compared to these functional groups, \sptfull has a slightly higher prediction error for carboxylic acid, nitrates and water as solutes and a noticeably higher prediction error for water as solvent, where the average error is \num{0.5}. Aliphatic hydrocarbons solved in water are the group combinations with the highest mean absolute error of \num{1.3}. However, due to the comparably high absolute $\ln \gamma$ of aliphatic hydrocarbons/water mixtures, a higher mean absolute error is expected. For water-containing systems, SPT-NRTL still significantly outperforms UNIFAC with an average mean absolute error of 0.50 on both \valedge and \valint  ($\mathrm{MAE}_\mathrm{max}=1.3$), while UNIFAC has an average mean absolute error across all groups of $\mathrm{MAE} = 1.3$ for water as solvent ($\mathrm{MAE}_\mathrm{max}=1.8$). 

Apart from water as solutes and solvent, \sptfull has the highest average mean absolute error for systems with carboxylic acids as solutes. While the average mean absolute error of the other functional groups used as solutes is between \numrange{0.1}{0.25}, the error of carboxylic acids used as solute is at \num{0.4} for \valint and \valedge and at \num{0.5} for \valext, close to the average mean absolute error of systems with water used as solute. Nevertheless, this error is lower than the prediction error or UNFIAC for the same systems at \num{0.5}.
Systems with carboxylic acids used as solvent have a lower average mean absolute error of \num{0.2} in \sptfull, similar to most other molecular families used as solvent. The deviation for carboxylic acids as solute might result from the low number of carboxylic acids (11 carboxylic acids in 103 unique mixtures) existing as solutes in the training data.

Overall, the analysis of the prediction error depending on the functional groups highlights the high accuracy of \sptfull compared to UNIFAC, with an average mean absolute error of \numrange{0.1}{0.2} for most of the functional group combinations. For these functional group combinations, the high prediction accuracy seems to approach a level of quality that corresponds to laboratory experiments.  

\subsection{Application of SPT-NRTL for VLE calculations}

\begin{figure}[b]
    \centering
    \captionsetup[subfigure]{oneside,margin={1cm,0cm}}
    \subfloat[Water/ethanol mixture]{
    \includegraphics[width=0.45\textwidth]{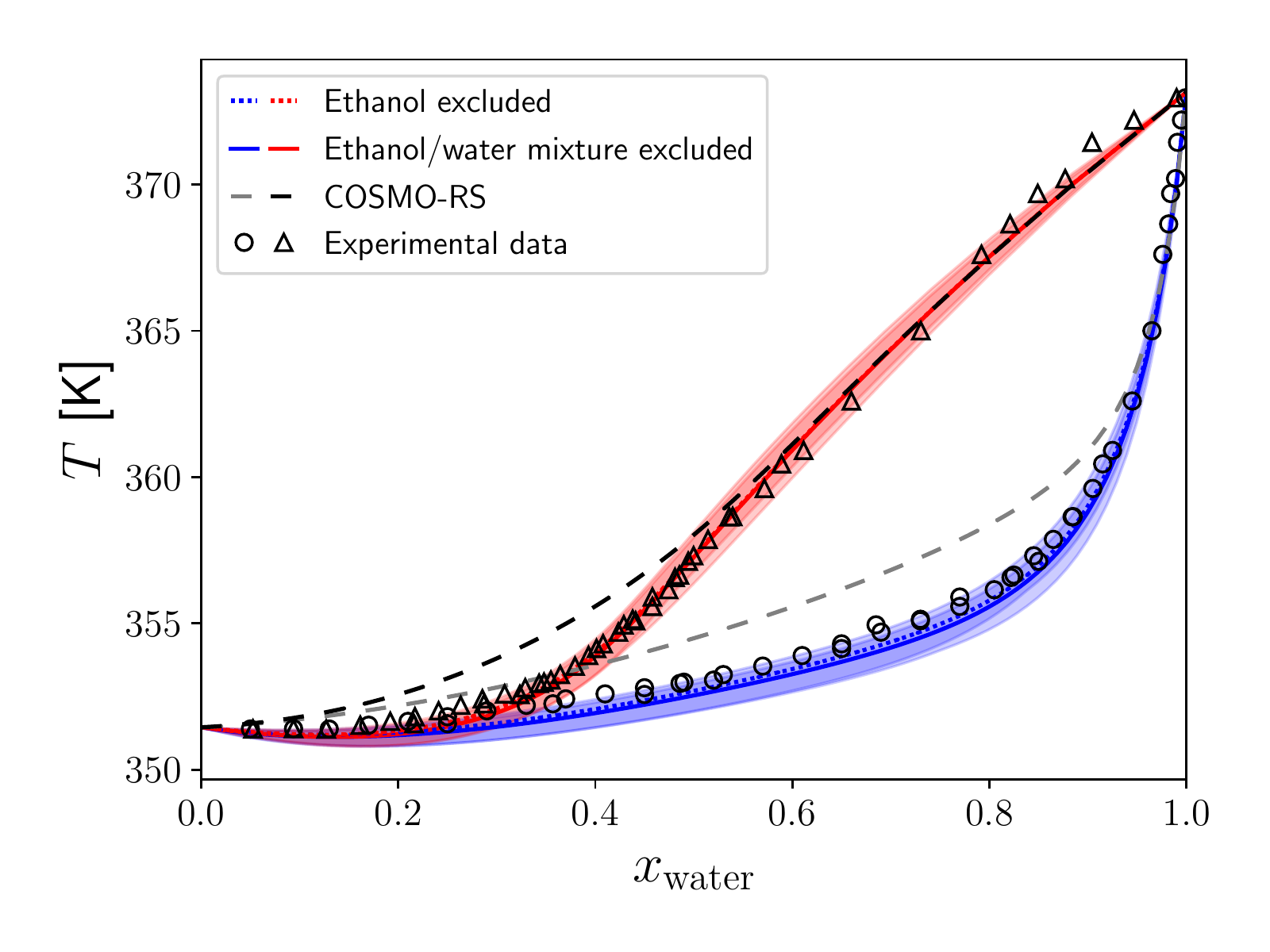}%
    \label{fig:VLEh2o}
    }
    \centering
    \captionsetup[subfigure]{oneside,margin={1cm,0cm}}
    \subfloat[Chloroform/n-hexane mixture]{
    \includegraphics[width=0.45\textwidth]{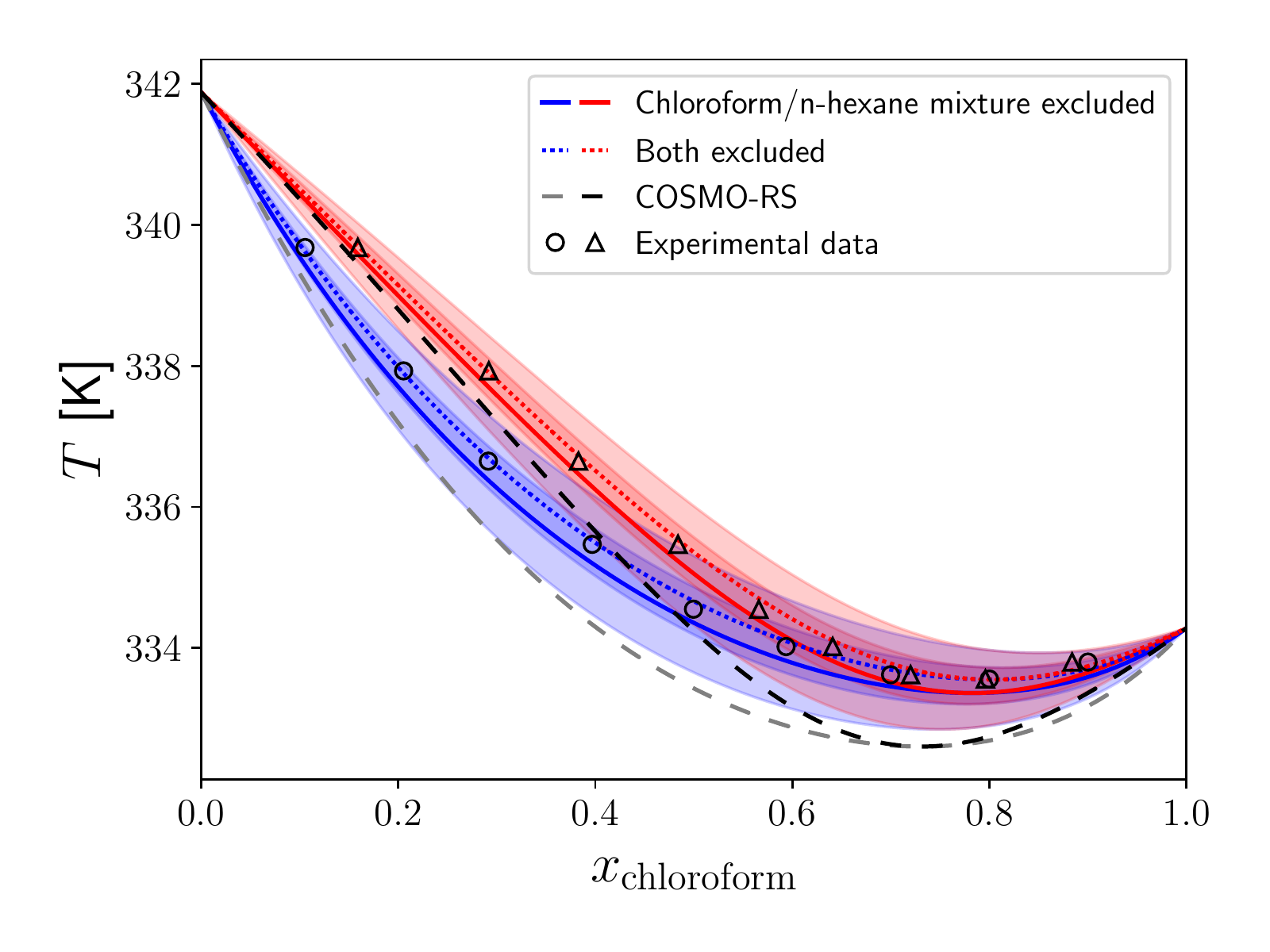}%
    \label{fig:VLEchloro}
    }
    \caption{Vapour-liquid equilibria for the water/ethanol and chloroform/n-hexane mixtures at \SI{1.013}{\bar}. The VLE is calculated using 20 SPT-NRTL models for each training set. The lines shows the mean prediction across the 20 models and the shaded area the minimum and maximum deviation. The black/grey line shows the prediction of the VLE using NRTL parameters fitted to COSMO-RS}
    \label{fig:VLE}
\end{figure}

In many applications in chemical engineering, NRTL-parameters are used to calculate phase equilibria like vapor-liquid equilibria (VLE). To asses the ability of \sptfull to calculate vapor-liquid equilibria, we use the NRTL parameters calculated by \sptfull for two selected mixtures.

We calculate the VLE for the mixtures water-ethanol and chloroform/n-hexane at \SI{1.013}{\bar}, both of which have an azeotrope (Figure \ref{fig:VLE}). To assess the capabilities of SPT-NRTL, we trained multiple models that exclude from the training data 1) the water/ethanol mixtures (interpolation) 2) all ethanol mixtures (edge extrapolation), 3) the chloroform/n-hexane mixtures (interpolation), and 4) all chloroform and all n-hexane mixtures (extrapolation). Due to the special treatment of water, we do not exclude water entirely from the fine-tuning. These models enable assessing the impact of unknown components on the VLE prediction.

When using the same training data, SPT will provide different slightly different outputs for each individual training run due to its stochastic nature. Thus, we trained \num{20} models for each of the four training sets and predicted the VLE with each of them.

For the water/ethanol mixture, experimental data are used from \citet{Jones.1943,Hughes.1952,Bloom.1961,Kojima.1968,StabnikovV.N.MatyushevB.Z..1972} and for the chloroform/n-hexane mixture, experimental data are used from \citet{Kudryavtseva.1963}, all data is provided by \citet{Jaubert.2020}, who established a high-quality reference database. To calculate vapour pressures, Antoine parameters from the NIST database are used \citep{Linstrom.1997}. Furthermore, we calculated NRTL parameters with COSMOtherm 2020 using TZVPD-FINE parameterization as benchmark for the prediction.

For both mixtures, water/ethanol (Figure~\ref{fig:VLEh2o}) and chloroform/n-hexane (Figure~\ref{fig:VLEchloro}), SPT-NRTL  is able to correctly predict the occurrence and position of the azeotrope with a high degree of accuracy, vastly outperforming COSMO-RS. For both the ethanol/\allowbreak water and the chloroform/\allowbreak n-hexane mixture, the uncertainty of the prediction shifts the dew/bubble-point line by about \SI{1}{K}. This \SI{1}{K} shift has a larger effect on the prediction of the chloroform/\allowbreak n-hexane mixture, as the absolute difference between boiling points is only \SI{7.6}{K} instead of the \SI{21.7}{K} for the water/ethanol mixture. 

These two examples highlight the high accuracy of the \sptfull model in calculating VLEs of mixtures contained in the training data and predict VLEs of mixtures not contained in the training data, even for mixtures with azeotropes. To lower the barrier to using SPT-NRTL, we calculated NRTL parameters for \num{100000000} mixtures with more than \num{10000} unique molecules with SPT-NRTL, which are available online (see Supporting Information Section 4). 
Due to the high accuracy of \sptfull for mixtures contained in the training data, the NRTL parameters of mixtures contained in the training data should be of high quality. For mixtures not contained in the training data (interpolation, and (edge-)extrapolation), the NRTL parameters should still capture important phenomena. However, a more systematic analysis of the ability of SPT-NRTL to correctly identify VLEs has to be performed in future work. 

\section{Conclusions}

The availability of property data remains a bottleneck in the development of new chemical processes. In particular, binary property data is scarce due to a large number of possible molecule combinations. In this work, we introduced SPT-NRTL, a transformer-based machine learning model to predict concentration-dependent thermodynamically consistent binary activity coefficients and their corresponding NRTL parameters. SPT-NRTL outperforms UNIFAC; particularly, at low concentrations, the average prediction error is halved.

An analysis of the prediction error depending on molecular families shows that SPT-NRTL is able to predict the binary activity coefficients for most functional groups with a mean absolute error between \num{0.1} and \num{0.2} in $\ln \gamma$, though water and carboxylic acids remain challenging with mean absolute errors increased to \num{0.4} and \num{0.5} in $\ln \gamma$. Nevertheless, even for water and carboxylic acids, the prediction quality is vastly improved compared to UNIFAC.

We demonstrate for two mixtures that SPT-NRTL is able to capture VLE behavior to near experimental accuracy. Thus, the provided \num{100000000} NRTL parameter sets calculated using SPT-NRTL should offer good quality data for many mixtures relevant to chemical engineering applications.

Beyond the specific use in the prediction of NRTL parameters, SPT-NRTL demonstrates the integration of physical equations, well-known in chemical engineering, into machine learning models for property prediction. Thereby, nearly any equation used for physical property calculation currently relying on fitted parameters could be transformed into a predictive model if sufficiently training data is available.

\section*{Acknowledgments}
This study was created as part of NCCR Catalysis (grant number 180544) a National Centre of Competence in Research funded by the Swiss National Science Foundation.

\section*{Author Contributions}
Benedikt Winter: Conceptualization, Data curation, Formal Analysis, Investigation, Methodology, Visualization, Writing – original draft, Writing – review \& editing.\\
Clemens Winter: Conceptualization, Methodology, Resources.\\
Timm Esper: Data Curation. \\
Johannes Schilling: Writing –Review \& Editing, Conceptualization, Methodology, Supervision.\\
Andr\'e Bardow: Writing - Review \& Editing, Conceptualization, Methodology, Supervision, Resources, Funding acquisition.\\

\bibliographystyle{abbrvnat}
\bibliography{lit}

\end{document}